\documentclass{article}
\usepackage[utf8]{inputenc}
\usepackage{amssymb}
\usepackage{amsmath}
\usepackage{amsfonts}
\usepackage{mathtools}
\usepackage{bbm}
\usepackage{xcolor}
\usepackage{tensor}
\usepackage{booktabs}
\usepackage{geometry}

\usepackage[inline]{enumitem}
\title{Projectively non-singular horizons in Kerr-NUT-de Sitter spacetimes}

\author{Jerzy Lewandowski \thanks{Jerzy.Lewandowski@fuw.edu.pl} \and Maciej Ossowski \thanks{Maciej.Ossowski@fuw.edu.pl}}

\date{%
\small
    Faculty of Physics, University of Warsaw,\\
ul. Pasteura 5, 02-093 Warsaw, Poland\\[2ex]%
    \today
}

\begin{document}

\maketitle

\begin{abstract}
It was recently discovered that Killing horizons in the generic Kerr-NUT-(anti) de Sitter spacetimes are projectively singular, i.e. their spaces of the null generators have singular geometry.
Only if the cosmological constant takes the special value determined by the Kerr and NUT parameters, and the radius of the horizon, then the corresponding horizon does not suffer that problem.
In the current paper, the projectively non-singular horizons are investigated.
They are found to be cosmological and non-extremal.
Every projectively non-singular horizon can be used to define a global completion of the Kerr-NUT-de Sitter spacetime it is contained in.
The resulting spacetime extends from $\mathcal{I}^-$ to $\mathcal{I}^+$, has the topology of $\mathbb{R}\times S_3$ and is smooth except for a possible Kerr-like singularity.
\end{abstract} 

\tableofcontents
\section{Introduction}
Our interest in the Kerr-NUT-(anti) de Sitter spacetimes comes from the theory of Petrov type D horizons, that are embeddable in spacetimes satisfying the $\Lambda$-vacuum Einstein equations \cite{geometryhorizonsAshtekar_2002,MechanicsIHPhysRevD.64.044016,DOBKOWSKIRYLKO2018415}.
They are given as solutions to a certain equation imposed on geometric data defined on a $2$-dimensional manifold.
The equation exhibits similar properties to those known in the global black hole theory as no rigidity \cite{Lewandowski_2006}, no hair \cite{localnohairPhysRevD.98.024008}, and uniqueness \cite{Lewandowski_2003} theorems.
The equation is also valid for horizons which do not admit a global spacelike cross-section because theirs null generators have the Hopf bundle structure.
A $4$-dimensional family of those horizons have been found \cite{localnohairPhysRevD.98.024008}.
Looking for their embeddings in the $4$-dimensional family of the Kerr-NUT-(anti) de Sitter spacetimes we discovered that, for a general (and generic) spacetime in this family, every Killing horizon is projectively singular.
We define the horizon to be projectively singular, if the $2$-dimensional metric induced on the space of the null generators of the horizon is singular.
In case of the Kerr-NUT-(anti) de Sitter horizons the induced metric suffers an irremovable discontinuity in at least one of the poles.
Only a $3$-dimensional subfamily of the Kerr-NUT-de Sitter (KNdS) spacetimes admits Killing horizons with a non-singular space of the null generators.
This is the family of the Kerr-NUT-de Sitter spacetimes we consider in the current paper.
Each of them satisfies the constraint that
\begin{equation}
\label{eq:thecond}  
 \Lambda =  \frac{3}{a^2+2l^2+2r_0^2},
\end{equation} 
where $a$ is the Kerr parameter, $l$ is the NUT parameter, and $r_0$ is the coordinate radius of the horizon. 
We investigate the location of the projectively non-singular horizon among the other horizons and check whether it can be extremal or not.
In a neighbourhood of the horizon, we study the geometry of the orbits of the corresponding Killing vector field.
Spacetimes with a non-zero NUT parameter have a conical singularity.
In the $\Lambda=0$ case that problem is known to be soluble by the Misner's glueing \cite{misner}.
Our decomposition of the metric tensor into a non-singular geometry of the orbit space and other geometric structures easily leads to a generalisation of the Misner's glueing.
This generalisation is valid for our Kerr-NUT- de Sitter spacetimes and extends from the past scri to the future one.

This paper is a direct continuation of our previous work \cite{LO}.

\section{Kerr-NUT-(anti) de Sitter spacetimes}
We consider the Kerr-NUT family of spacetimes with a cosmological constant parametrised by four real parameters $(m,a,l,\Lambda)$ and defined by the metric tensor \cite{gp2006,Griffiths_2007}
\begin{equation}
\label{eq:KNdS-metric}
   g=-\frac{\mathcal{Q}}{\Sigma}(dt-A d\phi)^2  +\frac{\Sigma}{\mathcal{Q}}dr^2 
+\frac{\Sigma}{P}d\theta^2+\frac{P}{\Sigma}\sin^2\theta(adt-\rho d\phi)^2,
\end{equation}
where
\begin{equation}\label{A,Q,...}
    \begin{split}
        \Sigma&=r^2+(l+a\cos\theta)^2,\\
        A&=a\sin^2\theta+4l\sin^2\tfrac{1}{2}\theta, \\
        \rho&=r^2+(l+a)^2=\Sigma+a A,\\
        \mathcal{Q}&=(a^2-l^2)-2m r+r^2-\Lambda\big((a^2-l^2)l^2+(\tfrac{1}{3}a^2+2l^2)r^2+\tfrac{1}{3}r^4\big),\\
        P&=1+\frac{4}{3}\Lambda a l  \cos\theta+\frac{\Lambda}{3}a^2\cos^2\theta.
    \end{split}
\end{equation}
Each of the metric tensors satisfies the vacuum Einstein equations with cosmological constant $\Lambda$
\begin{equation} 
R_{\mu\nu}-\tfrac{1}{2}R g_{\mu\nu} + \Lambda g_{\mu\nu} = 0.
\end{equation}
The parameters have their standard interpretation: $m$ is the mass of the black hole, $a$ is the Kerr parameter responsible for the black hole rotation, finally $l$ is the NUT parameter, whose topological relevance is discussed below.
The Kerr-NUT-(anti-) de Sitter family of solution, especially in relation to Kerr-de Sitter families, have been widely studied \cite{Mars2013,Mars2016,Mars2016b}.
The metric has expected limits to the Kerr solution when $l=0=\Lambda$ and to the Taub-NUT solution when $a=0=\Lambda$.
Allowing $\Lambda\not=0$ corresponds to either Kerr-(anti) de Sitter or Taub-NUT-(anti) de Sitter family of solutions and changes considerably the asymptotic behaviour at $r\rightarrow \infty$ \cite{Millerdoi:10.1063/1.1666343}.
The metric has several sources of potential singularities familiar for black hole spacetimes as vanishing of the functions $\mathcal{Q}$ and $\Sigma$, corresponding to black hole horizons and a curvature singularity, respectively \cite{griffiths_podolsky_2009}.
For that reason in the coordinate system $(t,r,\theta,\phi)$ the range of the variable $r$ is not entire $\mathbb{R}$, but an open interval.
For the purpose of the point we want to make below let us denote it by $I$.
Of course we will later use an Eddington-Finkelstein-like extension of that coordinate system, in order to get rid of the restriction to an interval.
A new possibility is vanishing of the function $P$ possible for some values of $\Lambda, l$ and $a$ \cite{LO}.
We assume such parameters that
\begin{equation*}
    P(\theta)>0, \ \ \ {\rm for\  all} \ \ \ \theta\in [0,\pi] .
\end{equation*}
The non-zero parameter $l$ introduces a notable topological change.
What is peculiar about this case is the singularity of the part of the differential $1$-form $A d\phi$, namely of
\begin{equation*}
    4l\sin^2\tfrac{1}{2}\theta d\phi.
\end{equation*}
It is discontinues at the pole $p_\pi$ such that $\theta=\pi$, when considered on a sphere parametrised by $(\theta,\phi)$.
The singularity can be cured by the transformation proposed by Misner \cite{misner}, namely 
\begin{equation*}
    t=t' + 4l\phi.
\end{equation*}
That amounts to the transformation 
\begin{equation}
\label{eq:A_rho_trans}
    A\mapsto A':=a\sin^2\theta-4l\cos^2\tfrac{1}{2}\theta, \quad {\rm and} \quad  \rho\mapsto \rho' :=r^2+(l-a)^2=\Sigma+aA' 
\end{equation}
 in (\ref{eq:KNdS-metric}).
 Now $A'$ is well defined at $p_\pi$, however it is singular at the pole $p_0$ such that $\theta=0$.
 Another price we pay is the condition that now $(t,r,\theta,\phi)$ and $(t+8l\pi,r,\theta,\phi) $ label the same point in spacetime.
 The result is two manifolds (generalised charts): $S^1\times  I\times \left(S^2\setminus \{p_\pi\}\right)$ parametrised by $(t,r,\theta,\phi)$ and $S^1\times I \times \left(S^2\setminus \{p_0\}\right)$ parametrised by $(t',r',\theta',\phi')$, with the transformation law 
$$ t=t' + 4l\phi',\ \ \ r=r',\ \ \  \theta=\theta',\ \ \  \phi=\phi' ,\ \ \ \ \ \ \ \ \ \ \text{for}\   \theta,\theta'\not=0,\pi .$$
Notice, that the topology of the resulting glued spacetime is $I\times S^3$.
There is another problem, though.
The angular part of the spacetime metric tensor (\ref{eq:KNdS-metric})
\begin{equation*}
    \frac{\Sigma}{P}d\theta^2+\frac{P}{\Sigma}\sin^2\theta \rho^2 d\phi^2
\end{equation*}
is discontinues at $\theta=0$, unless we introduce a rescaled, proper, angular coordinate $P(0)\phi$ ranging the interval $[0,2\pi)$.
On the other hand, for the primed chart valid at $\theta=\pi$, the corresponding condition at $p_\pi$ is that $P(\pi)\phi'$ is the proper angle variables running from $0$ to $2\pi$.
Since $\phi =\phi'$ where the charts overlaps, the consistency condition becomes
\begin{equation*}
    P(0) = P(\pi).
\end{equation*}
It is satisfied for $\Lambda=0$ or $l=0$ or $a=0$.
Otherwise in the general case, the metric tensor has an irremovable conical singularity on at least one of the $2$-surfaces, correspodning to $\theta=0$ or $\theta=\pi$. 

A conclusion is that we can not trust the common  meaning of the variables used in (\ref{eq:KNdS-metric}), for example surfaces $r=$const, $t=$const are not continues $2$-dimensional sections of the spacetime.
Therefore, we propose to investigate structures geometrically well defined and use them as our tools in the study of the Kerr-NUT-(anti) de Sitter spacetimes.
A geometrically well defined $3$-dimensional surface in a given spacetime (\ref{eq:KNdS-metric}) is one of the (maximally four) Killing horizons.
Its geometry is degenerate, but we can consider the $2$-dimensional space of the null generators.
It is endowed with the natural geometry - is it singular or not?
In the latter case, can the singularity be removed?
The extension of the space of the null generators of a Killing horizon is the $3$-dimensional space of the orbits of the corresponding Killing vector field, also naturally endowed with a metric tensor as long as the Killing vector is not null.
We can again ask, is it non-singular?
Exactly those questions were addressed in \cite{LO} and several surprising results were found.
New results are coming below in Sec. \ref{sec:newAboutHorizon} and \ref{sec:newOrbitSpace}.

\subsection{Horizons of non-singular null generator space in KNdS spacetimes}
The following section briefly summarises the results of our previous study of the Killing horizons and their neighbourhoods in the Kerr-NUT- (anti) de Sitter spacetimes \cite{LO}. 

In order to describe the horizons in the spacetime (\ref{eq:KNdS-metric}), we perform an Eddington-Finkelstein-like coordinate transformation
\begin{equation}
\label{eq:EddigntonCoords}
    dv:=dt+\frac{\rho}{\mathcal{Q}}dr,\quad
    d\Tilde{\phi}:=d\phi +\frac{a}{\mathcal{Q}}dr.
\end{equation}
The metric tensor (\ref{eq:KNdS-metric}) then becomes
\begin{equation}
\label{eq:metricThroughHorizon}
    ds^2=-\frac{\mathcal{Q}}{\Sigma}(dv-A d\Tilde{\phi})^2+2 dr (dv-A d\Tilde{\phi})+\frac{\Sigma}{P}d\theta^2+\frac{P}{\Sigma}\sin^2\theta(a dv-\rho d \Tilde{\phi})^2.
\end{equation}
As it is argued below in the Sec. \ref{sec:killingsHorizonsKNdS}, Killing horizons appear only when (and where) the function $\mathcal{Q}$ vanishes.
Let $H$ be a connected component of a such surface.
It is characterised by $r=r_0$, where $r_0$ is a root of the polynomial $Q$.
The metric tensor induced on $H$ is
\begin{equation}
\label{eq:q_H3}
q_H = \frac{\Sigma_0}{P}d\theta^2+\frac{P}{\Sigma_0}\sin^2\theta(a dv-\rho_0 d \Tilde{\phi})^2,
\end{equation}
where $\Sigma_0:=\Sigma(r_0)$ and $\rho_0:=\rho(r_0)$.
In three dimensions $q_H$ is degenerate and the degenerate direction is parallel to the Killing vector
\begin{equation}
\label{eq:xi}
\xi = \partial_v + \Omega_H \partial_{\tilde \phi}, \quad \Omega_H=  \frac{a}{\rho_0},
\end{equation}
which is simply $\xi = \partial_t + \Omega_H \partial_{\phi}$ in the more familiar $(t,r,\theta,\phi)$ coordinates.
In order to find the metric tensor induced by $q_H$ on the space of the null generators of $H$, we replace the coordinate $\Tilde{\phi}$ by a coordinate $\hat{\phi}$ constant along the null geodesics in $H$, that is 
\begin{equation}
\hat{\phi}:=-\Omega_H v+ \tilde{\phi}.
\end{equation}
Since 
$$\xi(\theta)=\xi(\hat{\phi})=0,$$
the coordinates $\theta$ and $\hat{\phi}$ parametrise the space of the null geodesics, and the metric tensor induced thereon is
\begin{equation}
\label{eq:horizon-2-metric}
{}^{(2)}q_H=\frac{\Sigma_0}{P}d\theta^2+\frac{P}{\Sigma_0}\sin^2\theta\rho_0^2 d \hat{\phi}^2.
\end{equation}

A curve of a constant $\theta=\theta_0$  parametrised by $\hat{\phi}\in[0,2\pi c)$, where $c$ is an unknown constant, is a circle around the pole $p_0$, as well as around the pole $p_\pi$. The corresponding geodesic radii are 
\begin{equation}
    R_0(\theta_0)=\int_0^{\theta_0}\sqrt{{}^{(2)}q_H{}_{\theta\theta}} d\theta, \quad R_\pi(\theta_0)=\int^\pi_{\theta_0}\sqrt{{}^{(2)}q_H{}_{\theta\theta}} d\theta. 
\end{equation}
The circumference of such circle is
\begin{equation}
    L(\theta_0)=\int_0^{2\pi c}\sqrt{q_H^{(2)}{}_{\hat{\phi}\hat{\phi}}} d\hat{\phi}.
\end{equation} 

By requiring
\begin{equation}
\label{eq:theconstraint}
\lim_{\theta_0\rightarrow 0}\frac{L(\theta_0)}{R_0(\theta_0)} = 2\pi = \lim_{\theta_0\rightarrow \pi}\frac{L(\theta_0)}{R_\pi(\theta_0)},
\end{equation} 
we obtaine the singularity removal condition
\begin{equation}
\label{eq:condition-for-diff-horizon}
    \frac{P(0)}{r_0^2+(l+a)^2}=\frac{P(\pi)}{r_0^2+(l-a)^2}, \quad c=1/P(0).
\end{equation}
For generic values of the parameters $m,a,\Lambda,l$ the condition (\ref{eq:condition-for-diff-horizon}) is not satisfied and the metric tensor (\ref{eq:horizon-2-metric}) has an irremovable  discontinuity  at one of the poles.
The singularity can be removed at both of the poles simultaneously if and only if: either $l=0$, or $a=0$, or  \cite{LO}:
 \begin{equation}
 \label{eq:constraint-Lambda}
  \Lambda =  \frac{3}{a^2+2l^2+2r_0^2},
  \end{equation}
and both $a$ and  $l$ are non-zero.
The case  $r_0=0$ requires a separate calculation, we discuss it in Sec. \ref{sec:r0=0} and show that this possibility can be excluded.
Hence, the parameters $a,l$ and $r_0$ freely take all the real values such that $r_0\not=0$.
The parameter $m$ is determined by the vanishing of $\mathcal{Q}$ at $r=r_0$ and the constraint (\ref{eq:constraint-Lambda}),
namely
\begin{equation}
\label{eq:mass-constraint}  
m =\frac{a^4 - 2 a^2 l^2 + l^4 + 2 a^2 r_0^2 - 6 l^2 r_0^2 + r_0^4}{2 a^2 r_0 + 4 l^2 r_0 + 4 r_0^3}.
\end{equation} 
The symmetry $(r_0,m)\mapsto (-r_0,-m)$ may be applied to ensure $m\ge 0$.   

The non-singular geometry of the space of the null generators is obtained via introducing a new rescaled coordinate $\varphi$ defined as
\begin{equation}
\label{eq:varphi}
 \varphi := \frac{\hat{\phi}}{P(0)},  
 \end{equation}
 and assuming it is cyclic with the period  $2\pi$. 

In conclusion, the horizons of the non-singular null generator space contained in the Kerr-NUT de Sitter spacetimes are given by $a,l,r_0$ freely taking non-zero values and $\Lambda,m$ determined by (\ref{eq:constraint-Lambda}, \ref{eq:mass-constraint}).
Replacing the $m$ parameter by $r_0$ has several drawbacks.
Firstly, it is immediate from (\ref{eq:constraint-Lambda}) that given a Kerr-NUT-de Sitter spacetime only one of the horizons, precisely that of the coordinate radius $r_0$, can have a non-singular null generator geometry.
Secondly, we do not know the location of the root $r_0$ among other roots of the polynomial $\mathcal{Q}$ reconstructed in such way that $r_0$ is a root and the parameters satisfy (\ref{eq:constraint-Lambda}, \ref{eq:mass-constraint}).
Thirdly, we do not know whether it is a simple root or double or of higher multiplicity. 
Two latter questions are answered by our new results presented   below in Sec. \ref{sec:newAboutHorizon}. 

\bigskip

The null generators of $H$ are orbits tangent to the Killing vector field $\xi$.
We also examined the space of the orbits of $\xi$ in neighbourhood of $H$ \cite{chrusciel}.
A technical assumption about $H$ we needed to make was that it is a non-extremal Killing horizon of $\xi$.
We show in Sec. \ref{sec:cosmologicalAndNonextremal} that the assumption was indeed justified.
Our approach relies on a decomposition - valid in a neighbourhood of $H$, except for points of $H$ itself - of the metric into: 
\begin{itemize}
\item the lapse function 
\begin{equation}\label{lapse}
    \xi^\mu\xi_\mu \not=0
\end{equation}

\item the rotation-connection 1-form
\begin{equation}\label{omega}
 \omega_\mu := \frac{\xi_\mu }{\xi^\nu\xi_\nu} 
\end{equation}

\item and the metric $q$ on the space of the orbits, that pulled back to the spacetime is
\begin{equation}\label{qcov}
 q_{\mu\nu} := g_{\mu\nu} - {\xi^\alpha \xi_\alpha} \omega_\mu \omega_\nu. 
\end{equation}
\end{itemize}
We parametrised the space of the orbits of the Killing vector $\xi$ by the functions $r$, $\theta$ and 
\begin{equation*}
 \hat{\phi}:=-\Omega_H t + \phi,
\end{equation*}
defined in spacetime (\ref{eq:KNdS-metric}) and constant along its orbits. 
The orbit space metric tensor  $q$  takes the following form:
\begin{equation}
\label{eq:q2}
q=  \frac{\Sigma}{\mathcal{Q}}dr^2 + \frac{\Sigma}{P}d\theta^2+\frac{P \mathcal{Q}\sin^2(x^2) \rho_0^2\Sigma}{\mathcal{Q}\Sigma_0^2- P \sin^2(x^2) a^2((x^1)^2-r_0^2)^2} d\hat{\phi}^2.  
\end{equation}
For a generic spacetime the above metric tensor has an irremovable conical singularity at the half axis $\theta=0$ or $\theta=\pi$.
However, we have found that the necessary and sufficient removability condition is again  
 (\ref{eq:constraint-Lambda}) with $r_0$ standing for the coordinate radius of the horizon $H$.
 Again, the required rescaling of $\hat{\phi}$ is that of (\ref{eq:varphi}).
 Upon those conditions the singularity disappears at the both half-axis and for all the values of $r$ as long as $\xi^\mu\xi_\mu\not=0$. 

With the non-singular metric tensor $q$ on the orbit space, the rotation-connection $\omega$ and non-vanishing lapse function $\xi^\mu\xi_\mu$ we proposed a new gluing scheme, alternative to the one presented in Section 2, that provides a non-singular neighbourhood for the horizon $H$.
Not knowing the location of the horizon $H$ among the other horizons though, we were able to implement that construction only locally \cite{LO}.
In the current paper we show that the non-singular spacetime built around the horizon $H$ can be extended to infinity.

\section{Properties of the non-singular Killing horizons}
\label{sec:newAboutHorizon}
In the first subsection we make sure that every Killing horizon in the Kerr-NUT-(anti) de Sitter spacetime (\ref{eq:KNdS-metric}) is defined by the vanishing of the function $\mathcal{Q}$.
For that purpose we adopt a method used for other black hole spacetimes in \cite{chrusciel}.
In the second subsection we find the position of the root $r_0$  corresponding to the non-singular Killing horizon with respect to the other roots of $\mathcal{Q}$.
Next, we show $r_0$ is a single root, hence the corresponding  horizon is  non-extremal.
In the last, short subsection we discuss the case $r_0=0$ and show that the corresponding Killing horizon has a curvature singularity \cite{griffiths_podolsky_2009}. 

\subsection{Killing horizons in KNdS spacetimes}
\label{sec:killingsHorizonsKNdS}
Let $H$ be a Killing horizon of the metric tensor (\ref{eq:KNdS-metric}).
It is developed by a Killing vector of a general form
\begin{equation}
\xi = B\partial_t + C \partial_{\phi}, \ \ \ \ \ \ B,C=\rm const,
\end{equation}
because  the  vector space of the Killing vectors fields  is spanned by  $\partial_t$ and $\partial_{\phi}$.   

The first observation is that $H$ contains two dimensional  orbits of the isometries generated by the Killing vectors $\partial_t$ and $\partial_{\phi}$.
It follows from the fact that the horizon $H$ is defined by the two equations:
 \begin{align}
g(B\partial_t + C \partial_{\phi}, B\partial_t + C \partial_{\phi}) &= 0,\\ 
 g(B\partial_t+ C \partial_{\phi}, B\partial_t + C \partial_{\phi})^{,\mu} \, g(B\partial_t+ C \partial_{\phi}, B\partial_t + C \partial_{\phi})_{,\mu}\ &=
 0. 
\end{align} 
Since each of these conditions is annihilated by each of the vector fields $\partial_t$ and $\partial_{\phi}$,  both the vector fields are tangent to $H$.

Now, the point is that at $H$ the orbit metric tensor, i.e. metric tensor restricted to the distribution spanned by $\partial_t$ and $\partial_{\phi}$, is degenerate.
The orbit metric tensor reads:
\begin{equation}
 -\frac{\mathcal{Q}}{\Sigma}(dt-A d\phi)^2+\frac{P}{\Sigma}\sin^2\theta(a dt-\rho d {\phi})^2.
\end{equation}  
The $1$-forms $dt-A d\phi$ and $a dt-\rho d {\phi}$ are linearly dependent if and only if $\rho - aA = 0$. However we have
 \begin{equation}
\rho - aA  = \Sigma,
\end{equation}
and thus such points correspond to an irremovable singularity at $\Sigma=0$.
Hence the metric degeneracy condition is   
\begin{equation}
\mathcal{Q}\sin^2\theta=0.
\end{equation} 
The vanishing of $\sin\theta$ corresponds to a polar coordinate singularity, so the only remaining possibility is that at the horizon we must have
\begin{equation}
\mathcal{Q}|_H=0.
\end{equation} 
The function $\mathcal{Q}(r)$ has isolated roots, therefore on each connected component of $H$, the coordinate $r$ equals exactly one of them  
\begin{equation}
\label{eq:r0}
r|_H=r_0.
\end{equation}  

The conclusion is that,  as long as $\mathcal{Q}\not=0$, the metric tensor (\ref{eq:KNdS-metric}) does not admit any Killing horizons. 

\subsection{Cosmological character and non-extremality of the non-singular horizons}
\label{sec:cosmologicalAndNonextremal}
In this subsection we find the location of the non-singular horizon.
Technically, it amounts to examining possible positions of the root $r_0$ among the other roots of the polynomial $\mathcal{Q}$, assuming that $\Lambda$ is given by (\ref{eq:constraint-Lambda}) and $m$ is that of (\ref{eq:mass-constraint}).  

Before analysing the root structure of the polynomial $\mathcal{Q}$ it is convenient to introduce a new variable $y$, such that
\begin{equation}
y=\tfrac{r}{r_0}-1.
\end{equation} 
Then we can introduce a more convenient form of the polynomial $\mathcal{Q}$
\begin{equation}
\Tilde{\mathcal{Q}}(y):=\mathcal{Q}(r= r_0(1+y))=-\frac{y \left(a^4-2 a^2 \left(l^2-\text{r0}^2\right)+l^4+2
   l^2 \text{r0}^2 (2 y+1)+\text{r0}^4 \left(y^3+4 y^2+4
   y+1\right)\right)}{a^2+2 \left(l^2+\text{r0}^2\right)}.
\end{equation}
The radius $r_0$ may be either negative or positive.
In both cases, $r_0$ now corresponds to $y=0$, while 
$y>0$ means that $r_0$ is between $r(y)$ and $0$, i.e. $y>0 \iff |r(y)|>|r_0| $.
  
Now we use the Descartes rule of signs \cite{descartes2001discourse}.
It introduces an index equal to the number of changes of the sign of the polynomial's coefficients.
The index is an upper bound on the number of positive roots of the polynomial.
The actual number of the positive roots may be equal to the index or smaller by an even positive number, the difference corresponding to the complex roots.

The coefficients of the transformed polynomial $\Tilde{\mathcal{Q}}$ are (up to the positive factor of $a^2+2(l^2+r_0^2$)):
\begin{equation}
\label{eq:coefficients_Q}
    \begin{split}
        \Tilde{\mathcal{Q}}_0=&\ 0,\\
        \Tilde{\mathcal{Q}}_1=&-a^4+2a^2(l^2-r^2_0)-(l^2+r_0^2)^2,\\
        \Tilde{\mathcal{Q}}_2=&-4r^2_0(a^2+l^2),\\
        \Tilde{\mathcal{Q}}_3=&-4r^4_0,\\
        \Tilde{\mathcal{Q}}_4=&-r^4_0.
    \end{split}
\end{equation}
It can be checked that $\Tilde{\mathcal{Q}}_1<0$.
Hence, the coefficients of $\Tilde{\mathcal{Q}}$ do not change the sign.
In the consequence, there are no roots greater than $y=0$.
This statement translated to the $r$ variable is that $r_0$ is always an extremal root of the polynomial $\mathcal{Q}$ - maximal if it positive or minimal if it is negative.
This means that a non-singular horizon in a Kerr-NUT-de Sitter spacetime is always the outermost, at the appropriate side of $r=0$.
Both possibilities still can occur even if we fix the mass parameter $m$ to be non-negative.

Another observation about a non-singular horizon $H$ we make is that $H$ is always non-extremal.
Indeed, let us go back to the polynomial $\mathcal{Q}$ such that $\Lambda$ satisfies (\ref{eq:constraint-Lambda}) and $m$ is determined by (\ref{eq:mass-constraint}). 
Dividing $\mathcal{Q}$ by $(r-r_0)$ and substituting $r=r_0$ we get
\begin{equation}
   -\frac{a^4-2 a^2\left(l^2-r_0^2\right)+\left(l^2+r_0^2\right)^2}
   {r_0 \left(a^2+2\left(l^2+r_0^2\right)\right)}.
\end{equation}

The numerator of the above is the same as $\Tilde{\mathcal{Q}}_1$ and never vanishes.
We conclude that $r=r_0$ cannot be a double root of $\mathcal{Q}$, necessarily making the non-singular horizon also a non-extremal one.

\subsection{The $r_0=0$ case}
\label{sec:r0=0}
Suppose that the polynomial $\mathcal{Q}$ has a root 
\begin{equation}\label{r=0}
r_0=0 
\end{equation}
and  $\Lambda$ satisfies the conical singularity removability condition (\ref{eq:constraint-Lambda}). 
As the lowest coefficient of $\mathcal{Q}$ is $(a^2-l^2)(1-l^2\Lambda)$, the condition (\ref{r=0}) implies
\begin{equation}\label{alt}
 a^2 = l^2, \ \ \ \ \ {\rm or}  \ \ \ \ \ 1-l^2\Lambda=0 .
 \end{equation}
In the case 
\begin{equation*}
    a=-l
\end{equation*}
we have at the horizon
\begin{equation*}
    \rho_0=0
\end{equation*}
and the  tensor $q_H^{(2)}$ has rank $1$.  On the other hand, in the case
\begin{equation*}
    a=l
\end{equation*}
$\Sigma$ vanishes at $\theta=\pi$ and the Gauss curvature of $q_H^{(2)}$ diverges at that point.
The last possibility is $\Lambda=1/l^2$.
Solving (\ref{eq:constraint-Lambda}) with this constraint requires that $a^2=l^2$ and leads back to the previous cases.
Moreover, other terms of the spacetime metric are also singular when $\Sigma$ vanishes. 
In conclusion, it is not possible for a non-singular Killing horizon to be located at $r=0$.
In this case the singularity becomes even worse than it can be for $r_0\not=0$.

\section{Non-singular space of Killing orbits and non-singular KNdS spacetime}
\label{sec:newOrbitSpace}
The starting point of this section is the metric tensor (\ref{eq:KNdS-metric}) that admits a Killing horizon $H$ of a non-singular space of the null generators.
The horizon is defined by a root $r_0$ of the polynomial $\mathcal{Q}$.
Our goals in this section are 
\begin{itemize}
\item  deriving the decomposition of the spacetime metric tensor into the structures perceived by the Killing observer in rest with respect to $H$ (see Section 2): the lapse function, the rotation-connection $1$-form and the geometry of the orbit space
\item using that decomposition for a systematic construction of two charts and a glueing recipe similar to the one presented in Section 2 that provides a conical singularity free spacetime containing the horizon $H$ 
\item extending the resulting spacetime in the variable $r$ from  $-\infty$ to $\infty$ (excluding the singularity at at possible zeros of the function $\Sigma$).  
\end{itemize}

Due to the symmetry 
\begin{equation}
\label{eq:-m}
    (r,m)\mapsto (-r,-m),
\end{equation} we can assume without losing arbitrariness that   
\begin{equation*}
     r_0 >0
\end{equation*}
(although $m$ may end up negative value).  Hence, according to the previous section, $r_0$ is the largest root of the polynomial $\mathcal{Q}$,
\begin{equation*}
    r_0=r_+.
\end{equation*}

We will use the coordinates $(v,r,\theta,\varphi)$ introduced above  (see the equalities (\ref{eq:EddigntonCoords}, \ref{eq:varphi})) because: 
\begin{enumerate*}[label=(\roman*)]
    \item they extend the spacetime (\ref{eq:KNdS-metric}) across the surfaces of vanishing of the function $\mathcal{Q}$ 
    \item the corresponding Killing vector $\xi$ takes a particularly simple form: $\xi = \partial_v$ and finally 
    \item when the condition (\ref{eq:constraint-Lambda}) holds the conical singularity of the space of the null generators of the horizon and  the conical singularity of the space of the orbits of $\xi$ is removed by the assumption that the period of the cyclic variable $\varphi$ is $2\pi$.
\end{enumerate*}

In terms of those coordinates the metric tensor (\ref{eq:KNdS-metric}) reads 
\begin{equation}
\label{eq:metricThroughHorizonNonsingular}
    g=-\frac{\mathcal{Q}}{\Sigma}\bigg(\frac{\Sigma_0}{\rho_0}dv-\frac{A}{P(0)} d\varphi\bigg)^2 
    +2 dr\bigg(\frac{\Sigma_0}{\rho_0}dv-\frac{A}{P(0)} d\varphi\bigg)
    +\frac{\Sigma}{P}d\theta^2
    +\frac{P}{\Sigma}\sin^2 \theta\Big(\frac{a}{\rho_0}\big(r^2-r_0^2) dv+\frac{\rho}{P(0)} d\varphi\Big)^2.
\end{equation}

\subsection{The region of  $\xi^\mu\xi_\mu\not=0$ }\label{sec:non-vanishing}
A necessary condition for a Killing observer is the non-vanishing of the lapse function (see Section 2), namely
\begin{equation*}
    \xi^\mu\xi_\mu \not=0.
\end{equation*}
Therefore, a particularly favourable circumstance is the following general property of {\it all} the KNdS  spacetimes ($\Lambda >0$): if $r_+$  is the largest root of the function $Q$ in (\ref{eq:KNdS-metric}), and $\xi$ is the Killing vector field tangent to the null generators of the corresponding Killing horizon then
\begin{equation}
\xi^\mu\xi_\mu >0 \quad {\rm for\ \ every} \quad r>r_+ .
\end{equation}
\bigskip
Indeed, a short calculation using the metric tensor (\ref{eq:metricThroughHorizonNonsingular}) gives
\begin{equation}
\label{eq:gxixi}
\xi^\mu\xi_\mu  = g_{vv} =  -\frac{\mathcal{Q}}{\Sigma}\bigg(\frac{\Sigma_0}{\rho_0}\bigg)^2+  \frac{P}{\Sigma}\sin^2 \theta\Big(\frac{a}{\rho_0}\big(r^2-r_0^2)\bigg)^2.
\end{equation}
The second term is manifestly non-negative.
The leading term of the polynomial $\mathcal{Q}$ is $-\frac{\Lambda}{3}r^4$ and makes the first term in (\ref{eq:gxixi}) positive  for every $r>r_+$. 

We also have some control over the sign of $\xi^\mu\xi_\mu$ for an interval $(r_0-\epsilon,r_0)$ provided  $\epsilon>0$ is sufficiently small. That is true since, as it is proved above, the root $r_0$ corresponding to a non-singular horizon must be a simple root.
In consequence, the first term in (\ref{eq:gxixi}) is linear in $r-r_0$.
This makes the second term negligible for sufficiently small $\epsilon>0$ as it is proportional to  $(r-r_0)^2$. 
The first term is negative as long, as $r$ continues to be bigger than the second largest root of $\mathcal{Q}$.   

Another immediate gain is determination that the signature of ${}^{(2)}q_H$ is $(+,+)$.
The $d\theta^2$ component is always positive due to the assumption $P>0$.
On the other hand $\mathcal{Q}<0$ implies that both the numerator and the denominator of the $d\phi^2$ component are negative.

\subsection{Decomposition of the spacetime metric}

Whenever in spacetime the Killing  vector $\xi$ is neither null nor zero, we can naturally decompose the metric tensor into terms of the lapse function (\ref{lapse}), the rotation-connection $\omega_\mu dx^\mu$ (\ref{omega}), and the pullback $q_{\mu\nu}dx^\mu dx^\nu$ (\ref{qcov}) of the orbit space metric \cite{chrusciel}.
The decompoition reads as
 \begin{equation}
 \label{eq:decomp}
 g =  \xi^\alpha\xi_\alpha\omega_\mu\omega_\nu dx^\mu dx^\nu + q.  
 \end{equation}
 In terms of our coordinates we have 
\begin{equation}
\label{eq:split}
    g=g_{vv}(dv + \omega_i dx^i)^2 + q_{ij}dx^idx^j, \quad x^i=r,\theta,\varphi, 
\end{equation}
where
\begin{equation}
\begin{split}
    q&=\frac{\Sigma_0^2\Sigma}{\mathcal{Q}\Sigma^2_0-P\sin^2\theta a^2(r^2-r_0^2)^2} dr^2+\frac{P a \Sigma^2(r^2-r_0^2)\rho_0\sin^2\theta}{P(0)\Sigma(\mathcal{Q}\Sigma_0^2-P\sin^2\theta a^2(r^2-r_0^2)^2)}2 dr d\varphi\\
    &+\frac{\Sigma}{P}d\theta^2+\frac{P\mathcal{Q}\sin^2\theta\rho_0^2\Sigma}{\mathcal{Q}\Sigma^2_0-P\sin^2\theta a^2(r^2-r_0^2)^2}\frac{1}{P^2(0)}d\varphi^2\label{q},\\
    \omega_r&=-\frac{\Sigma\Sigma_0\rho_0}{\mathcal{Q}\Sigma^2_0-P\sin^2\theta a^2(r^2-r_0^2)^2},\\
    \omega_\varphi&=-\frac{\rho_0}{P(0)}\frac{A\Sigma_0\mathcal{Q}+P\sin^2\theta a\rho(r^2-r_0^2)}{\mathcal{Q}\Sigma^2_0-P\sin^2\theta a^2(r^2-r_0^2)^2},\\
    \omega_\theta&=0.\\
\end{split}
\end{equation}
That decomposition is well defined as long as the lapse function $\xi^\mu\xi_\mu=g_{vv}\neq0$. 

\subsection{The  part orthogonal to $\xi$}
The part $q$ satisfies
\begin{equation}
\xi^\mu q_{\mu\nu} = 0 = {\cal L}_\xi q
\end{equation}
and is the pullback of the geometry induced on  the orbit space.
According to the conclusions of Section \ref{sec:non-vanishing},  the denominators present in (\ref{q}) do not vanish for $r\in (r_0-\epsilon,r_0)\cup (r_0,\infty) $.
For a generic KNdS spacetime, there is an irremovable singularity of the geometry $q$ produced by the terms proportional to $d\theta^2$ and $d\varphi^2$ along at least one of the axis $\theta=0$ or $\theta=\pi$.
However, if the geometry of the space of the null generators of the Killing horizon developed by $\xi$ is non-singular, that is when the condition (\ref{eq:constraint-Lambda}) is satisfied, then the singularity of $q$ is also removable simultaneously along the both axis at all the values of $r$ such that $\xi^\mu\xi_\mu\not=0$ \cite{LO}.
The term of $q$ proportional to $dr d\varphi$ is also a smooth tensor for all the values of $\theta$ including $0$ and $\pi$ provided $r\in (r_0-\epsilon,r_0)\cup (r_0,\infty) $.
The same is true for the $dr^2$ term.
In conclusion, $q$ is smoothly defined on all the space of the orbits  of the Killing vector $\xi$ diffeomorphic to    
\begin{equation*}
    \left(\left(r_0-\epsilon,r_0\right)\cup \left(r_0,\infty\right)\right) \times S^2.
\end{equation*}

\subsection{The rotation-connection part and a second chart} 
The troublesome part of  $g_{vv}(dv + \omega_i dx^i)^2$, which is parallel - parallel to $\xi_\mu dx^\mu$, is the $1$-form  
\begin{equation}
\omega = dv + \omega_\varphi d\varphi + \omega_r dr .
\end{equation}
Again, it follows from the arguments presented in Section \ref{sec:non-vanishing} that the denominators present in (\ref{q}) do not vanish as long as  $r\in  (r_0-\epsilon,r_0)\cup (r_0,\infty)$.
However the term $\omega_\varphi d\varphi $ has the discontinuity at $\theta=\pi$. The obstacle is  the non-zero value of $\omega_\varphi(r,\pi)$, namely
\begin{equation}
 \omega_\varphi(r,\theta=\pi)=\frac{-4l}{P(\pi)}.
\end{equation}
This problem is cured  by introducing another chart $(v',r',\theta',\varphi')$ valid for $\theta'\not=0$ (however invalid for $\theta'=0$)  and related to $(v,r,\theta,\varphi)$  by the transformation
\begin{equation}
\label{eq:t'}
v=v' + \frac{4l}{P(\pi)}\varphi', \ \ \ r=r', \ \ \ \theta=\theta', \ \ \ \varphi=\varphi'\ \ \ \ \ {\rm for}, \ \ \ \theta,\theta'\not= 0,\pi . 
\end{equation}
 Then we have
  \begin{equation}
  \label{eq:omegatrans}
dv + \omega_\varphi d\varphi + \omega_r dr = dv' + \omega'_{\varphi'}d\varphi' +  \omega'_{r'} dr', \ \ \ \ \ \ \omega'_{\varphi'}(r',\theta')d\varphi':= \left(\omega_\varphi(r',\theta') +\frac{4l}{P(\pi)}\right)d\varphi', \ \ \ \ \omega'_{r'}(r',\theta') := \omega_r(r',\theta'),
\end{equation} 
and the $1$-form $\omega$ is well defined in all the domain of $(t',r',\theta',\varphi')$ including $\theta'=\pi$. 
 
Since we want the transformation (\ref{eq:t'}) to be well defined for all values of $\varphi$ we have to take into the account, that  $\varphi$ and $\varphi+2\pi$ correspond to the same point in the spacetime, for given values of $v,r,\theta$.
Hence, the same has to be true about $v$ and $v+\frac{4l}{P(\pi)}2\pi$, given values of $r,\theta,\varphi$.
That means that $v$ has also to be a cyclic variable and its period has to be a multiple of $\frac{4l}{P(\pi)}2\pi$.
In consequence, the topology of the domain of the (generalised) chart $(v,r,\theta,\varphi)$ is
\begin{equation}
 S^1\times\left((r_0-\epsilon,r_0)\cup (r_0,\infty)\right)\times (S^2\setminus \{p_\pi\}).
\end{equation}
The analogous arguments show the domain of the primed chart $(v',r',\theta',\varphi')$ is 
\begin{equation}
 S^1\times\left((r_0-\epsilon,r_0)\cup (r_0,\infty)\right)\times (S^2\setminus \{p_0\}),
\end{equation}
and $v'$ has the same period as $v$.   

The $1$-forms $dv + \omega_\varphi d\varphi + \omega_r dr$ and $dv' + \omega'_{\varphi'}d\varphi' +  \omega'_{r'} dr'$ consistently  define an everywhere smooth (and even analytic) differential $1$-form $\omega$. 

\subsection{Glueing the  metric tensor}
The lapse function $g_{vv}$  and the orthogonal part $q$ of the metric tensor are invariant with respect to the transformation (\ref{eq:t'}), hence they pass unchanged to  the primed chart 
\begin{equation}
g_{v'v'}(r',\theta') :=  g_{vv}(r',\theta'), \ \ \ \ q=q'_{ab}(r',\theta')dx'^adx'^b := q_{ab}(r',\theta')dx'^a dx'^b. 
\end{equation}

In this way, all of the metric tensor $g$ has been  extended  onto the primed chart domain such that it is smooth both at $\theta=0$ and at $\theta'=\pi$.
All the mechanism of the extension consists in (\ref{eq:omegatrans}) and freedom from the conical singularity of the orbit space geometry $q$.   

\subsection{Extending the glueing to all $r\in \mathbb{R}$.} 
Now we go back to the  metric tensor $g$ written as in (\ref{eq:metricThroughHorizonNonsingular}),
\begin{equation}\label{eq:metricThroughHorizonNonsingular2}
    g=-\frac{\mathcal{Q}}{\Sigma}\bigg(\frac{\Sigma_0}{\rho_0}dv-\frac{A}{P(0)} d\varphi\bigg)^2 
    +2 dr\bigg(\frac{\Sigma_0}{\rho_0}dv-\frac{A}{P(0)} d\varphi\bigg)
    +\frac{\Sigma}{P}d\theta^2
    +\frac{P}{\Sigma}\sin^2 \theta\Big(\frac{a}{\rho_0}\big(r^2-r_0^2) dv+\frac{\rho}{P(0)} d\varphi\Big)^2,
\end{equation}
and consider it in the following domain  of the variables $(v,r,\theta,\varphi)$
\begin{equation}
 S^1\times\mathbb{R}\times (S^2\setminus \{p_\pi\}).
\end{equation}
This is the first chart. We also consider another chart defined in
\begin{equation}
 S^1\times\mathbb{R}\times (S^2\setminus \{p_0\}),
\end{equation}
endowed with variables $(v',r',\theta',\varphi')$, and related with the first one by the transformation (\ref{eq:t'}).
In the second chart the metric tensor becomes
\begin{equation}\label{eq:metricThroughHorizonNonsingular'}
    g=-\frac{\mathcal{Q}}{\Sigma}\bigg(\frac{\Sigma_0}{\rho_0}dv'-\frac{A'}{P(\pi)} d\varphi'\bigg)^2 
    +2 dr'\bigg(\frac{\Sigma_0}{\rho_0}dv'-\frac{A'}{P(\pi)} d\varphi'\bigg)
    +\frac{\Sigma}{P}d\theta'^2
    +\frac{P}{\Sigma}\sin^2 \theta'\Big(\frac{a}{\rho_0}\big(r'^2-r_0^2) dv'+\frac{\rho'}{P(\pi)} d\varphi'\Big)^2
\end{equation}
where $A'$ and $\rho'$ are defined same as previously by (\ref{eq:A_rho_trans}) (up to trivial identities that $r=r'$ and $\theta=\theta'$)
\begin{equation}
A'(\theta') := a\sin^2\theta'-4l\cos^2\tfrac{1}{2}\theta',\ \ \ \ \ \ \ \ \ \ \rho'(r'):=r'^2 +(a-l)^2.
\end{equation}

It is not obvious at the first sight, but one can check by inspection, that in the both domains the metric tensor is well defined, smooth and even analytic.
This is owing to the conditions (\ref{eq:condition-for-diff-horizon}) which are equivalent to (\ref{eq:thecond}). 

\section{The resulting spacetime}
 The resulting topology of the spacetime is $\mathbb{R}\times S^3$. The topology of each of the surfaces 
$$r={\rm const} $$
is that of $S^3$.
In particular, the $3$-spheres corresponding to the roots of the function $\mathcal{Q}$ are Killing horizon.
The Killing vector field $\xi$ generates an action of the group O(2) that induces  the topological structure of the Hopf bundle 
\begin{equation*}
    S^3 \rightarrow S^2,
\end{equation*}
which can be naturally extended to 
\begin{equation*}
    \mathbb{R}\times S^3\rightarrow \mathbb{R}\times S^2.
\end{equation*}
For the horizon $H$ corresponding to $r_0$ the orbits of $\xi$ overlap with the null generators.
For other horizons though, this is not the case.
Also, the geometry of the space of the null generators is non-singular for $H$ only, not for the other horizons.  

The fate of the (would be) axial symmetry of spacetime requires some clarification.
The vector field $\partial_\varphi$ defined originally in the not primed chart, transforms to the primed chart as follows:
$$ \partial_\varphi = \partial_{\varphi'} - \frac{4l}{P(\pi)}\partial_{v'} .$$
While it vanishes at $\theta=0$, it fails to vanish at $\theta'=\pi$.
On the other hand, we have a similar, primed  symmetry generator
$$ \partial_{\varphi'} =  \partial_\varphi  + \frac{4l}{P(\pi)}\partial_{v}, $$
that vanishes at $\theta'=\pi$, however does not vanish at $\theta=0$. 
A democratic choice is
$$\Phi := \frac{1}{2}\left(\partial_\varphi + \partial_{\varphi'} \right)=  \partial_\varphi  + \frac{2l}{P(\pi)}\partial_{v} = \partial_{\varphi'}  - \frac{2l}{P(\pi)}\partial_{v'} $$
however that vector does not vanish anywhere.  

\section{Summary}
In this paper we continued our investigation of the Kerr-NUT- de Sitter spacetimes containing a non-singular Killing horizon, that is a Killing horizon of a non-singular geometry of the space of the null generators.
This subfamily of the KNdS spacetimes (\ref{eq:KNdS-metric}) can be freely parametrised by: the Kerr parameter $a$, the NUT parameter $l$ and the coordinate radius $r_0\not=0$ of the horizon, which is the value the coordinate $r$ assumes at the non-singular horizon.
The cosmological constant $\Lambda$ and the mass parameter $m$ featuring in (\ref{eq:KNdS-metric}) are determined given arbitrary values of $(a,l,r_0)$.
In the current paper we examined the polynomial $\mathcal{Q}$ determined by an arbitrary triple $(a,l,r_0)$.
We showed that the root $r_0$ is:
\begin{itemize}
\item  the largest/smallest root of  $\mathcal{Q}$ - the symmetry $(\ref{eq:-m})$ maps one into the other
\item  a simple root - hence the corresponding horizon is non-extremal.
 \end{itemize} 
Since the cosmological constant is strictly positive, the horizon is cosmological.  

The decomposition of the spacetime metric tensor into the geometric characteristics of the Killing observer corresponding to the horizon at $r=r_0$ (the lapse function, rotation-connection $1$-form, and the non-singular orbit space) suggested a simple gluing receipt that provided a conical singularity free neighbourhood of the horizon.
Moreover, the Eddington-Finkelstein-like extension of the corresponding spacetime provides a non-extendable spacetime, which is also non-singular except for possible zeros of the function $\Sigma$.
The resulting spacetime has the topology of $\mathbb{R}\times S^3$.
The conformal completion is bounded by the future and past scris, endowed with a conformal geometry defined in a non-singular manner on $S^3$.

\textit{Acknowledgements} 
This work was partially supported by the Polish National Science Centre
grants No. 2017/27/B/ST2/02806 and No. 2016/23/P/ST1/04195.

\bibliographystyle{unsrt}
\bibliography{bibliography}

\end{document}